\pdfoutput=1
\documentclass[apjl]{emulateapj}

\usepackage{epstopdf}
\usepackage{lineno}
\usepackage{xspace}
\usepackage{graphicx}
\usepackage{amssymb}
\usepackage{amsmath}
\usepackage{natbib}
\usepackage[colorlinks=true,linkcolor=blue,citecolor=blue,urlcolor=blue]{hyperref}

\newcommand{\nustar}{\textsl{NuSTAR}\xspace}

\shorttitle{Disk-Jet Coupling in 4C+74.26}
\shortauthors{Bhatta et al.}

\begin{document}

\title{ Signatures of the disk-jet coupling in the Broad-line Radio Quasar 4C+74.26}

\author{
G.~Bhatta\altaffilmark{1},
{\L}.~Stawarz\altaffilmark{1},
A.~Markowitz\altaffilmark{2,\,3},
K.~Balasubramaniam\altaffilmark{1},
S.~Zola\altaffilmark{1,\,4},
A.~A.~Zdziarski\altaffilmark{2},
M.~Jamrozy\altaffilmark{1},
M.~Ostrowski\altaffilmark{1},
A.~Kuzmicz\altaffilmark{1,5},
W.~Og{\l}oza\altaffilmark{4},
M.~Dr\'o\.zd\.z\altaffilmark{4},
M.~Siwak\altaffilmark{4},
D.~Kozie\l-Wierzbowska\altaffilmark{1},
B.~Debski\altaffilmark{1},
T.~Kundera\altaffilmark{1},
G.~Stachowski\altaffilmark{4},
J.~Machalski\altaffilmark{1},
 V.~S.~Paliya\altaffilmark{6},
 and D.~B.~Caton\altaffilmark{7}
}

\altaffiltext{1}{Astronomical Observatory of the Jagiellonian University, ul. Orla 171, 30-244 Krak\'ow, Poland}
\altaffiltext{2}{Nicolaus Copernicus Astronomical Center, Polish Academy of Sciences, Bartycka 18, PL-00-716 Warsaw, Poland}
\altaffiltext{3}{University of California, San Diego, Center for Astrophysics and Space Sciences, 9500 Gilman Drive, La Jolla, CA 92093-0424, USA}
\altaffiltext{4}{Mt.\ Suhora Observatory, Pedagogical University, ul.\ Podchorazych 2, 30-084 Krak\'ow, Poland}
\altaffiltext{5}{Center for Theoretical Physics, Polish Academy of Sciences, Al. Lotnik\'ow  32/46, 02-668 Warsaw, Poland}
\altaffiltext{6} {Department of Physics and Astronomy, Clemson University, Kinard Lab of Physics, Clemson, SC 29634-0978, USA}
\altaffiltext{7}{Dark Sky Observatory, Department of Physics and Astronomy, Appalachian State University, Boone, NC 28608, USA}
\email{email: {\tt gopalbhatta716@gmail.com}}

\begin{abstract}
Here we explore the disk-jet connection in the broad-line radio quasar 4C+74.26, utilizing the results of the multiwavelength monitoring of the source. The target is unique in that its radiative output at radio wavelengths is dominated by a moderately-beamed nuclear jet, at optical frequencies by the accretion disk, and in the hard X-ray range by the disk corona. Our analysis reveals a correlation (local and global significance of 96\% and 98\%, respectively) between the optical and radio bands, with the disk lagging behind the jet by $250 \pm 42$ days. We discuss the possible explanation for this, speculating that the observed disk and the jet flux changes are generated by magnetic fluctuations originating within the innermost parts of a truncated disk, and that the lag is related to a delayed radiative response of the disk when compared with the propagation timescale of magnetic perturbations along relativistic outflow. This scenario is supported by the re-analysis of the NuSTAR data, modelled in terms of a relativistic reflection from the disk illuminated by the coronal emission, which returns the inner disk radius $R_{\rm in}/R_{\rm ISCO} =35^{+40}_{-16}$. We discuss the global energetics in the system, arguing that while the accretion proceeds at the Eddington rate, with the accretion-related bolometric luminosity $L_{\rm bol} \sim 9 \times 10^{46}$\,erg\,s$^{-1}$\,$\sim 0.2 L_{\rm Edd}$, the jet total kinetic energy $L_\textrm{j} \sim 4 \times 10^{44}$\,erg\,s$^{-1}$, inferred from the dynamical modelling of the giant radio lobes in the source, constitutes only a small fraction of the available accretion power.
\end{abstract}


\section{Introduction}
\label{sec:intro}
 
A symbiosis between accretion disks and jets in active galactic nuclei (AGN) is one of the widely studied topics in high-energy astrophysics, crucial to our understanding of the AGN-related feedback in the evolution of galaxies and, in general, the structure formation in the Universe. The exact connection between the disk and the jet state transitions 
 --- a widely researched topic in the case of black-hole X-ray binaries \citep[e.g.][]{Fender2004} ---  in AGN has only been established so far in the broad line radio galaxies (BLRGs) 3C\,111 and 3C\,120, where the dips in the X-ray light curves of the accretion disk coronae are followed by the ejection of superluminal knots along pc-scale radio jets \citep[e.g.][]{Marscher2002,Chatterjee09,Chatterjee2011}. In addition, \citet{Tombesi2012,Tombesi2013,Tombesi2017} presented the evidence for the presence of mildly-relativistic disk outflows in both sources, possibly related to the ``X-ray dips/radio-outburst'' cycles, and powerful enough to affect the surrounding interstellar medium in a feedback loop, as well as to collimate the radio jets.  Any comparable characterization of the disk-jet connection in other radio galaxies, would require a long-term monitoring of particularly bright sources, for which the disk and the jet emission components are both \emph{accessible} observationally. This, in fact, turns out exceedingly difficult in either blazars or narrow-line radio galaxies (NLRGs): in the case of blazars the Doppler boosted non-thermal jet emission mostly outshines the radiative output of accretion disks and disk coronae, whereas in NLRGs, where the jet is aligned at large angles to the line of sight, one typically deals with a severe obscuration of the central engine by hot dusty tori. This makes BLRGs the best-suited targets for investigating the disk-jet coupling, as in BLRGs the jets (at least their inner segments) are oriented only at intermediate viewing angles, which are smaller than the opening angles of obscuring tori, so that the disks and their coronae may be observed directly.

\begin{figure*}[t!]
\includegraphics[angle =-90,width=\textwidth]{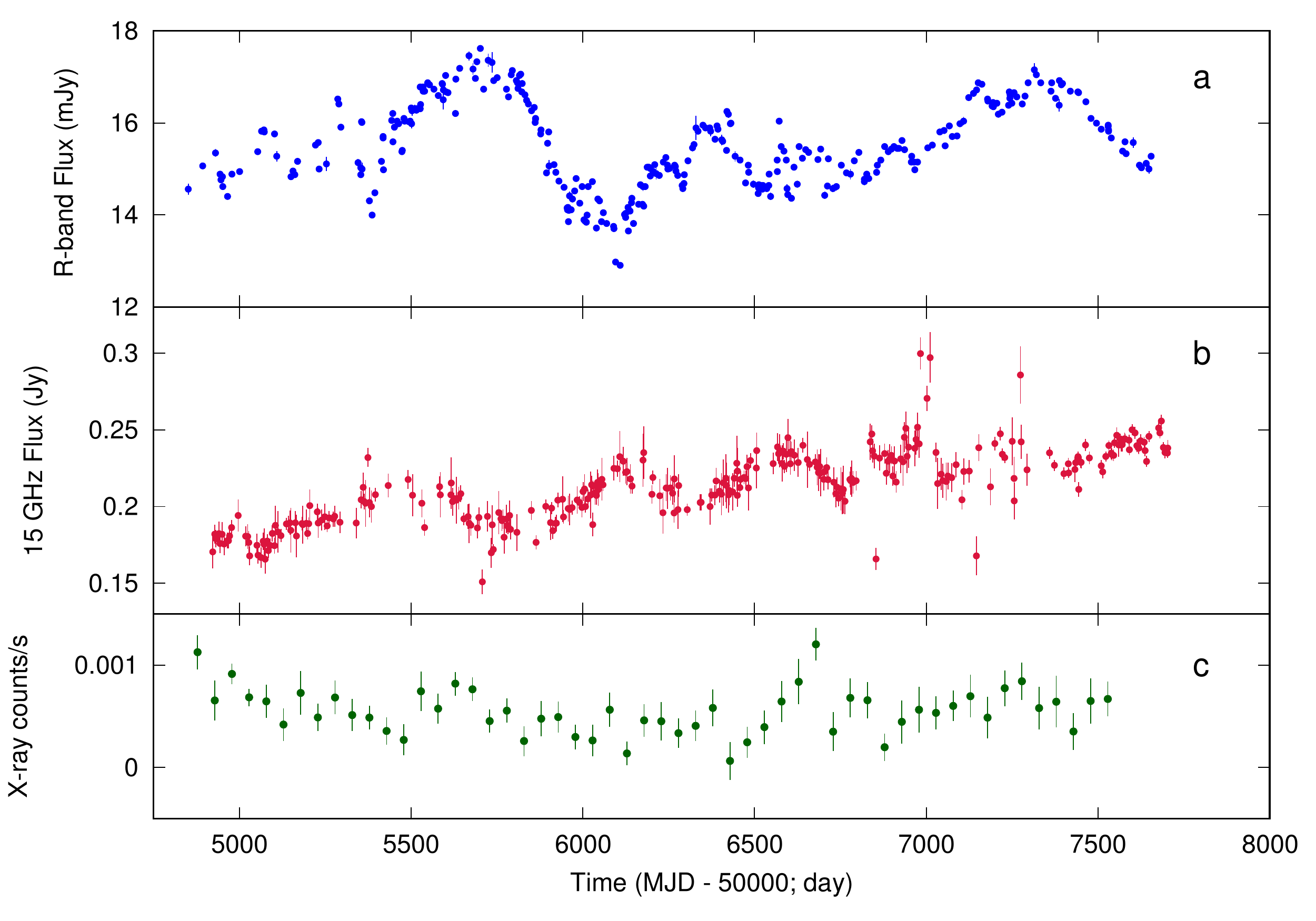}\\
\caption{Multiwavelength long-term light curves of 4C\,+74.26 spanning $\sim$\,eight years, and including the optical (R band) data mainly from the Suhora Observatory (upper panel\,a), the radio (15\,GHz) data from the OVRO monitoring program (middle panel\,b), and the hard X-ray ($14-50$\,keV) fluxes from the {\it Swift}-BAT Hard X-ray Transient Monitor program, weighted averaged in 50\,day bins (lower panel\,c). }
\label{LightCurves}
\end{figure*}

4C\,+74.26 (J2000.0 coordinates: R.A.=$\rm 20^{h}42^{m}37\fs3$ and Dec.= $+75\degr08\arcmin02\farcs5$), introduced by \citet{Riley1988}, is one of the largest known radio sources associated with an active nucleus classified spectroscopically as a broad-line radio galaxy/quasar. It displays giant radio lobes of the FR\,II type \citep{Fanaroff1974,Saripalli1996,Schoenmakers2001,Lara2001,Konar2004,Machalski2011}, with the total angular size of $\sim 10\arcmin$ at the redshift of $z=0.104$ \citep{Riley1988}, yielding a projected linear size of $\sim 1.01$\,Mpc for the modern cosmology. \citet {Pearson1992} first reported on the detection of the pc-scale radio jet in the source, and of the radio core with an inverted (i.e., self-absorbed) spectrum. Since no evidence for the counter-jet was found in the radio maps, and assuming that the jet-counterjet asymmetry is attributed solely to the Doppler beaming effect, an upper limit to the angle between the axis of the nuclear jet  and the line of sight can be estimated as $\leq 49\degr$. The jet can be followed up to hundreds-of-kpc scales in the  Very Large Array (VLA) radio maps \citep{Riley1990}, terminating in a strongly polarized and particularly X-ray-luminous hotspot \citep{Erlund2007,Erlund2010,Araudo15}. The total and the core luminosities of 4C\,+74.26 at 1.4\,GHz are $P_{\rm tot}= 10^{25.67}$\,W\,Hz$^{-1}$ and $P_{\rm core}= 10^{24.72}$\,W\,Hz$^{-1}$, respectively \citep{Kuzmicz2012}.

At X-ray frequencies, the source has been monitored by many instruments since 1993. In particular, Advanced Satellite for Cosmology and Astrophysics (\textit{ASCA}), ROentgen SATellite (\textit{ROSAT}), and \textit{BeppoSAX} spectra revealed a power-law component with the photon index $\Gamma_{\rm X} \sim 2$ and a neutral iron line with an equivalent width (EW) of $\sim 100$\,eV, Compton reflection, and indications of a warm absorber \citep[e.g.][]{Brinkmann1998,Sambruna1999,Reeves2000,Komossa2000,Hasenkopf2002}. \citet{Ballantyne05} reported on the analysis of the complex \textit{XMM-Newton} spectrum of the source, consisting of the emissions from ionised and neutral reflectors, cold and warm absorbers, and the broad neutral iron line. \citet{Larsson2008}, \citet{Fukazawa2011} and \citet{Patrick2012}, based on the \textit{Suzaku} data analysis, confirmed the presence of the iron line, but with an EW of only $< 100$\,eV. On the other hand, \citet{Gofford2013} and \citet{Tombesi2014} claimed the detection of the iron line at 6.5\,keV, suggestive of a mildly-ionized reflector instead of a neutral one, as well as a marginal detection of blue-shifted narrow absorption lines indicative of highly-ionized warm absorber. We note in this context that, in the optical range, \citet{Robinson1999} found the variations in the degree and the position angle of the polarized H${\alpha}$ line, implying high-speed ($\geq$ 0.01c) outflows along the 4C\,+74.26 jet. 

More recently, \citet{DiGesu2016}, using the {\it XMM-Newton} and {\it Chandra} grating spectroscopy, confirmed the presence of a warm absorber in the source with outflow velocities $\sim 0.01c$. Also, \citet{Lohfink2017}, based on the \nustar broad-band observations, confirmed the presence of the ionized reflection, and concluded that the disk is mildly recessed with an inner radius of $\sim$\,ten times the gravitational radius.

So far, 4C\,+74.26 has not been detected in high-energy $\gamma$-rays with the Large Area Telescope (LAT) onboard the {\it Fermi} satellite \citep{Kataoka2011}, unlike the two other aforementioned BLRGs 3C\,111 and 3C\,120 \citep[see][respectively]{Grandi12,Tanaka15}.

In this paper, we report on the systematical analysis of the optical, radio, and hard X-ray monitoring data for 4C\,+74.26 (see Section~\ref{sec:obs} below for the description of the instrument used, as well as of the corresponding data acquisition and data processing). Based on the discrete correlation function and the power spectral analysis, we found that the optical (disk) emission in the source lags behind the radio (jet) emission by about 250 days (see Section~\ref{sec:analysis}). We discuss the possible explanation for this lag relying on the truncated disk model (Section~\ref{sec:discussion}), as supported by the re-analysis of the \nustar data. 

Throughout the paper we assume the modern cosmology with $H_0 =  73$\,km\,s$^{-1}$\,Mpc$^{-1}$, $\Omega_{\rm m} = 0.27$, and $\Omega_{\Lambda} = 0.73$, so that the luminosity distance of the source is $d_L = 459$\,Mpc, and 1\,arcsec corresponds to 1.829\,kpc.

\section{Observations and Data Processing} 
\label{sec:obs}
 
\subsection{Monitoring data}
\label{sec:monitoring}

 Since 2006, we have been performing a long-term monitoring program of radio-loud AGN, initially focused at measuring the brightness of OJ\,287 in the optical band. Observations were carried out at the Mt. Suhora Observatory of the Pedagogical University in Krak\'ow, and at the Astronomical Observatory of the Jagiellonian University. Since 2009, we appended the list of targets with a sample of eight radio quasars described in \citet{Zola2015}. The measurements at Mt. Suhora were done with the 60-cm telescope and an Apogee CCD, mounted in the prime focus. A smaller, 50-cm telescope with an Andor CCD was used at the Jagiellonian University Observatory. Both telescopes are equipped with sets of wide band UBVRI filters purchased from the same manufacturer and made according to the \citet{Bessell1990} prescription. In addition, during the prolonged periods of cloudy weather at the two sites, data were also taken with the $14"$ robotic telescope at the Dark Sky Observatory, Appalachian State University, NC, USA. Since the sample of QSOs consists of objects with the brightness in the range between 12 and 16.5 magnitude (R filter), we decided to measure the targets in the R filter to maximize the S/N, due to the highest sensitivity of the back illuminated CCD chips at the red wavelengths.

We took several individual images of 4C\,+74.26 every few nights, along with the calibration images (bias, dark, and  flat field, usually on the sky). On several occasions, multicolor data (BVRI) were also taken but since they are scarce, we did not use them for the purpose of this paper. The reduction was done by a single person to ensure  results as uniform as possible; also, the same stars were used during the entire observing campaign and for all three sites as a comparison (GSC0103002595) and check (GSC0103002595) calibrators. We noticed no systematic shifts larger than 0.01 magnitude in the mean magnitude differences between comparison and check stars among the three sites over the entire period of monitoring. Therefore, all the measurements we took were simply combined.

The radio data for the target utilized in this paper, were taken from the 15\,GHz monitoring performed with the 40-m telescope at the Owens Valley Radio Observatory \citep[OVRO;][]{Richards2011}.

The 14--50 keV data were taken from the Hard X-ray Transient Monitor program\footnote{\url{https://swift.gsfc.nasa.gov/results/transients}} \citep{Krimm2013}, conducted with the Burst Alert Telescope \citep[BAT;][]{Barthelmy2005} onboard the {\it Swift} satellite \citep{Gehrels2004}. Due to a relatively low flux level of the source, resulting in large errors as well as occasionally negative count rates in daily bins, the daily data with zero data flag were selected and binned in 50-day bins using error weighted mean, and again in 100-day bin for the cross-correlation study. 

The resulting long-term light curves of 4C\,+74.26 in the optical, radio, and hard X-ray bands, spanning nearly eight years of the monitoring, are shown in Figure~\ref{LightCurves}.  Note a very different angular resolution of the instruments utilized here: while our optical telescopes are characterised by point spread functions (PSF) of about $\gtrsim 2''$, the 40\,m OVRO antenna at 15\,GHz has a beam size of $\gtrsim 1'$, and the {\it Swift}-BAT PSF is $\sim 17'$. Still, keeping in mind the timescales covered in our monitoring program (from weeks up to $\sim$\,eight years), any variability observed at optical, radio, and X-ray photon energy ranges, is expected to originate in the AGN itself, including the inner parts of the accretion disk, and the (sub)pc-scale jet.

\begin{figure}[t!]
\includegraphics[width=0.95\columnwidth]{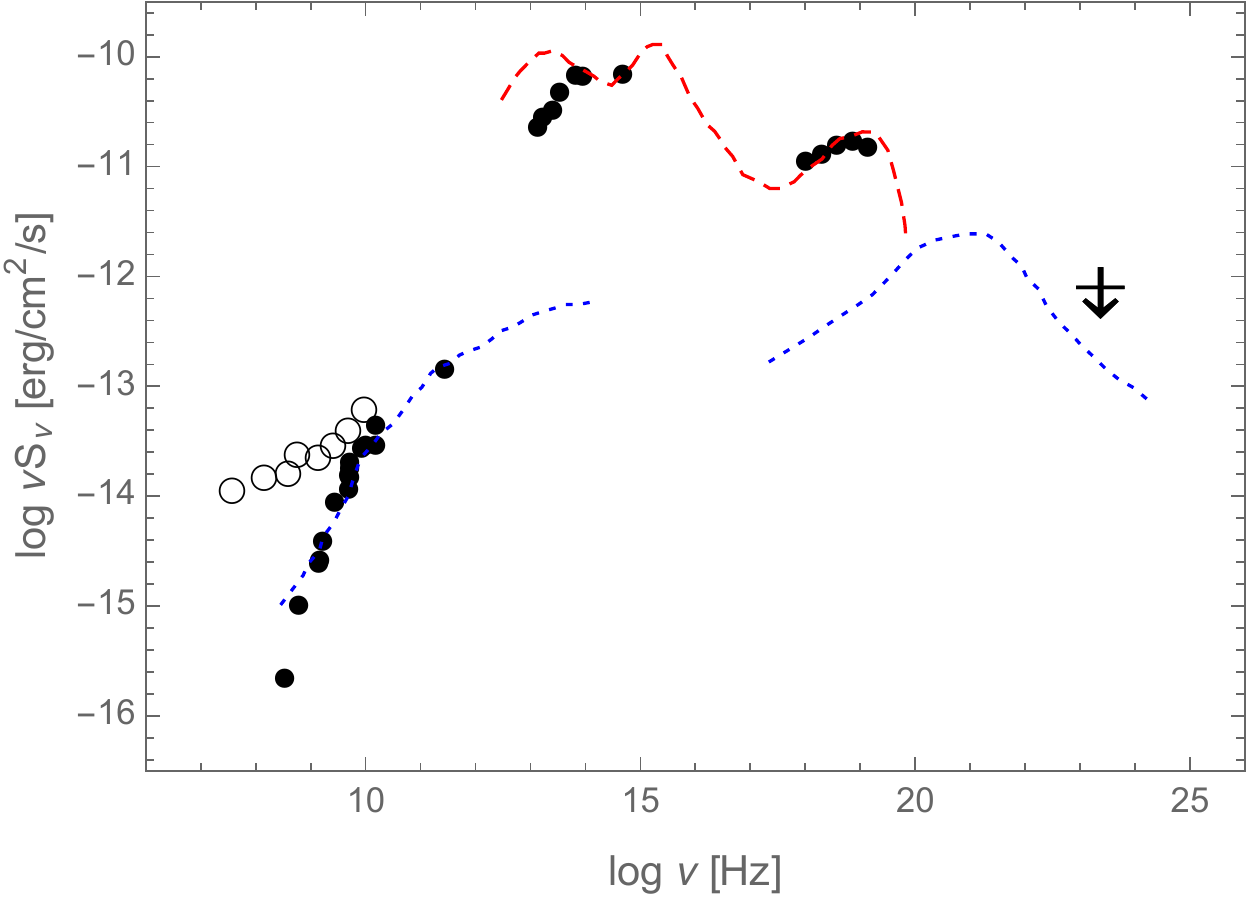}\\
\caption{Broad-band SED of 4C\,+74.26 (see Section~\ref{sec:SED} for the description of the multifrequency data included in the figure). The core fluxes are denoted by black filled circles, and the total radio fluxes by empty circles. We matched the SED datapoints with the normalized ``accretion-related'' quasar template from \citet[red long-dashed curve]{Koratkar99}, and the ``jet-related'' 3C\,273 template from \citet[blue short-dashed curve]{Soldi08}.
}
\label{SED}
\end{figure}

\subsection{\nustar Observations}
 \label{sec:nustar}

Nuclear Spectroscopic Telescope Array (\textsl{NuSTAR}), a space-bound instrument operating in the hard X-ray band (3--79 keV) with the spectral resolution of $\sim 1$\,keV, is perfectly suited for detailed spectral studies of accreting black holes. It consists of the two co-aligned focal plane modules, FPMA and FPMB, each with the field of view of $\sim 13$\,arcmins, and the half-power diameter of an image of a point source of $\sim 1$\,arcmin \citep[see][for more details]{Harrison2013}. The archival \nustar data for 4C\,+74.26 were processed using the \texttt{nupipeline} script from the \nustar Data Analysis Software (NuSTARDAS) package v.1.3.1. The cleaned event files were generated using the \nustar CALDB calibration files. The data were extracted from a region of 50\,arcsec radius, centered on the source, while the background was extracted from a 100\,arcsec radius region near the source location containing no other sources. The FPMA and FPMB light curves and the spectral files were obtained using the tasks \texttt{nuproducts}. The spectra were binned using \texttt{grppha} in order to have at least 30 counts per re-binned channel \citep[for further details see][]{Bhatta2018b}, and the spectral fitting was performed in \texttt{XSPEC}\footnote{\url{https://heasarc.gsfc.nasa.gov/xanadu/xspec}} version 12.10.0 \citep{Arnaud1996}.

\begin{table}[!th]
\centering
\caption{Multifrequency Discrete Correlation Analysis}
\begin{tabular}{lcclll} 
\hline
Bands & DCF value & Time lag$^{**}$ & Local & Global \\
 &  & (day) & sig. (\%)& sig. (\%)\\
\hline
Optical--Radio & 0.42$\pm$0.04 & $+ 250\pm42$  & 96& 98\\
X-ray--Optical & 0.36$\pm$0.11 & $-105\pm58$  & 89 & 87\\
X-ray--Radio & 0.24$\pm$0.10 & $+210\pm61$ & 77& 76\\
\hline
\end{tabular} 
\label{tab:DCF}
$^{**}$ Positive values indicate hard lags.
\end{table}

\subsection{Broadband Spectral Energy Distribution}
\label{sec:SED}
  
For inspecting the broad-band spectral energy distribution (SED) of the source 4C\,+74.26, in addition to the optical R-band, radio 15\,GHz, and the \nustar data described above, we have also gathered the multifrequency observations at other radio frequencies from various previous studies as listed in Table~1 of \citet[including the core and also the total fluxes]{Zola2015}, at infrared frequencies from \citet{Matsuta2012} and \citet{Ichikawa2017}, as well as the {\it Fermi}-LAT upper limits derived in \citet{Kataoka2011}. The resulting SED is presented in Figure~\ref{SED}.

\begin{figure}[t!]
\includegraphics[width=\columnwidth]{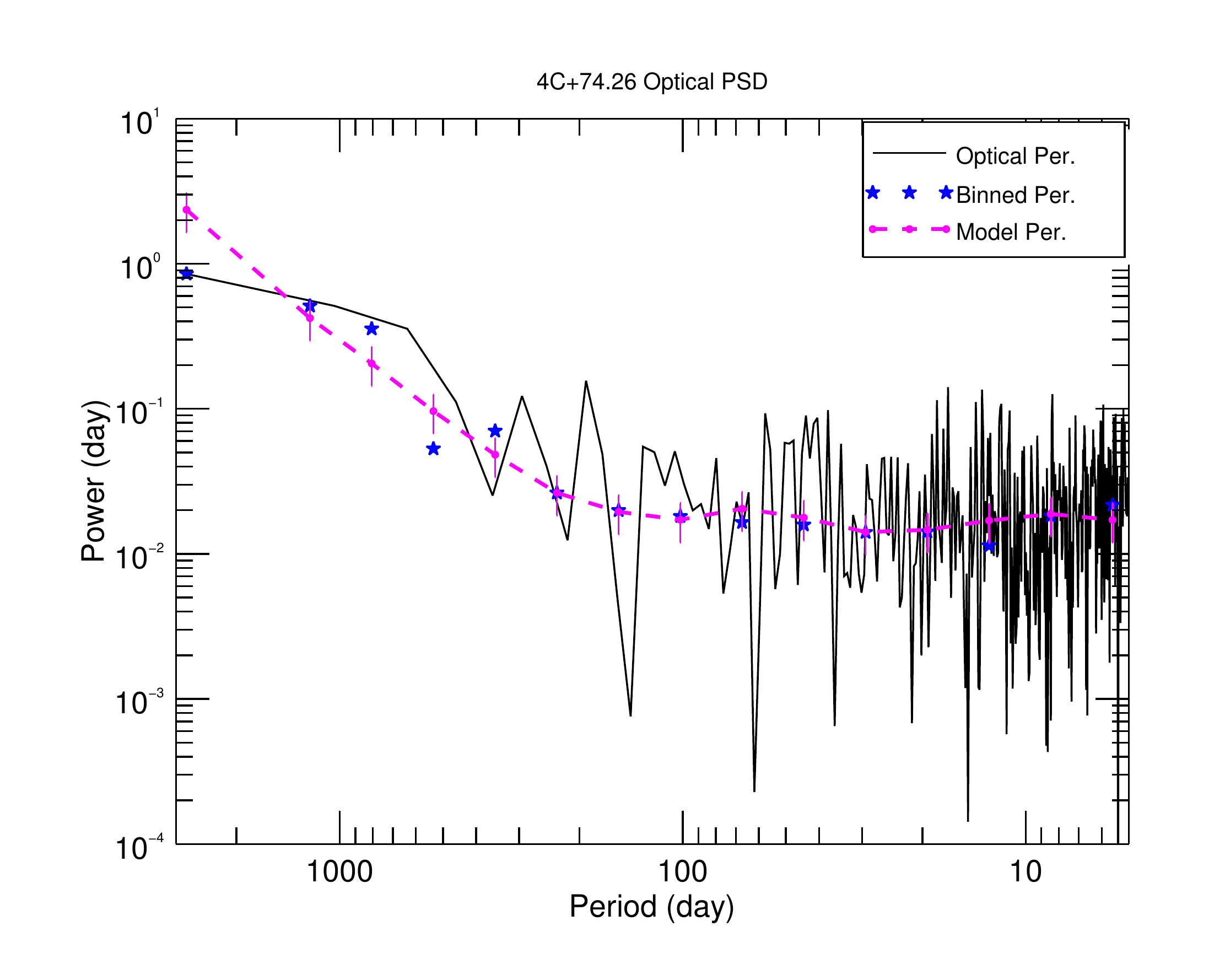}\\
\includegraphics[width=\columnwidth]{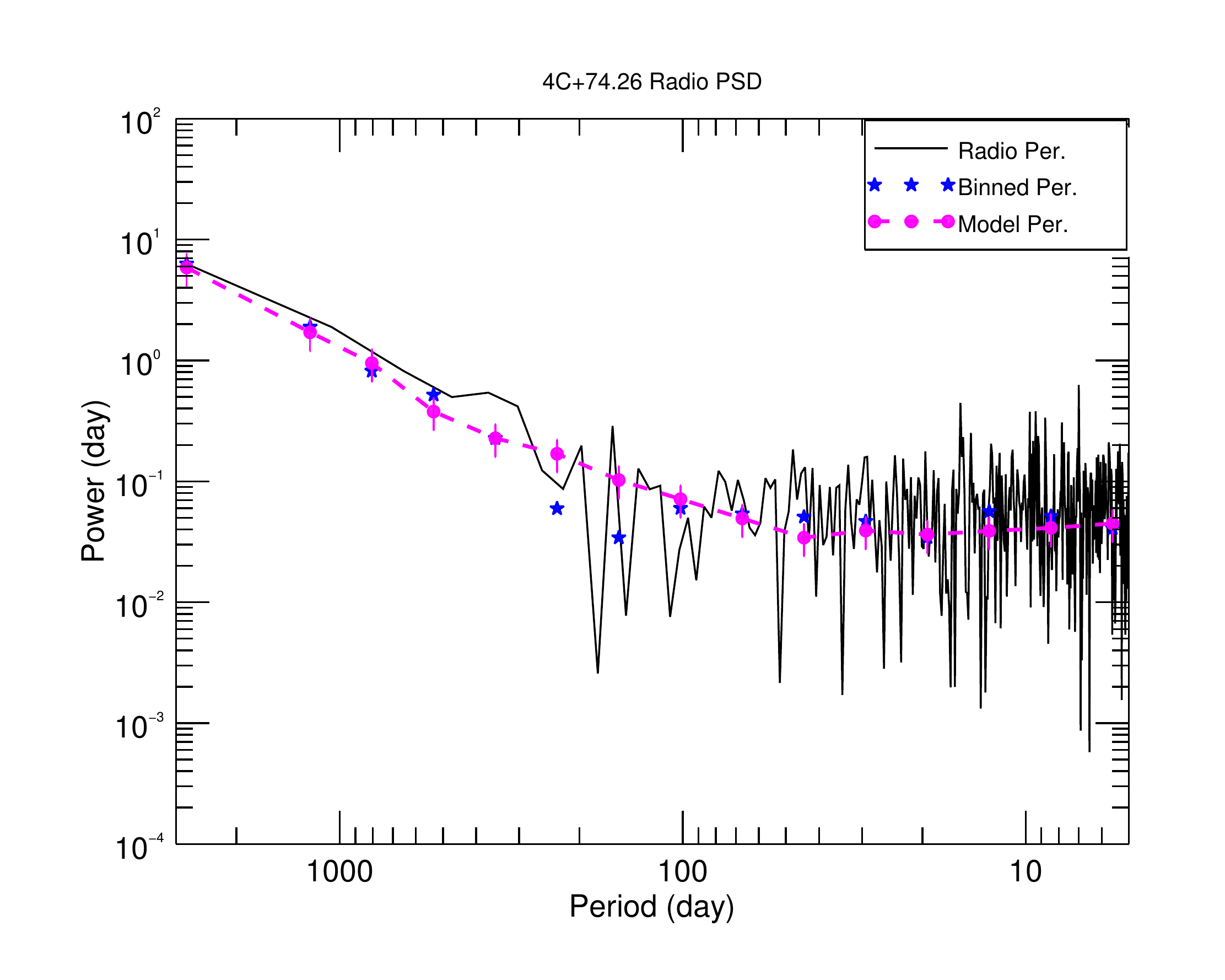}
\includegraphics[width=\columnwidth]{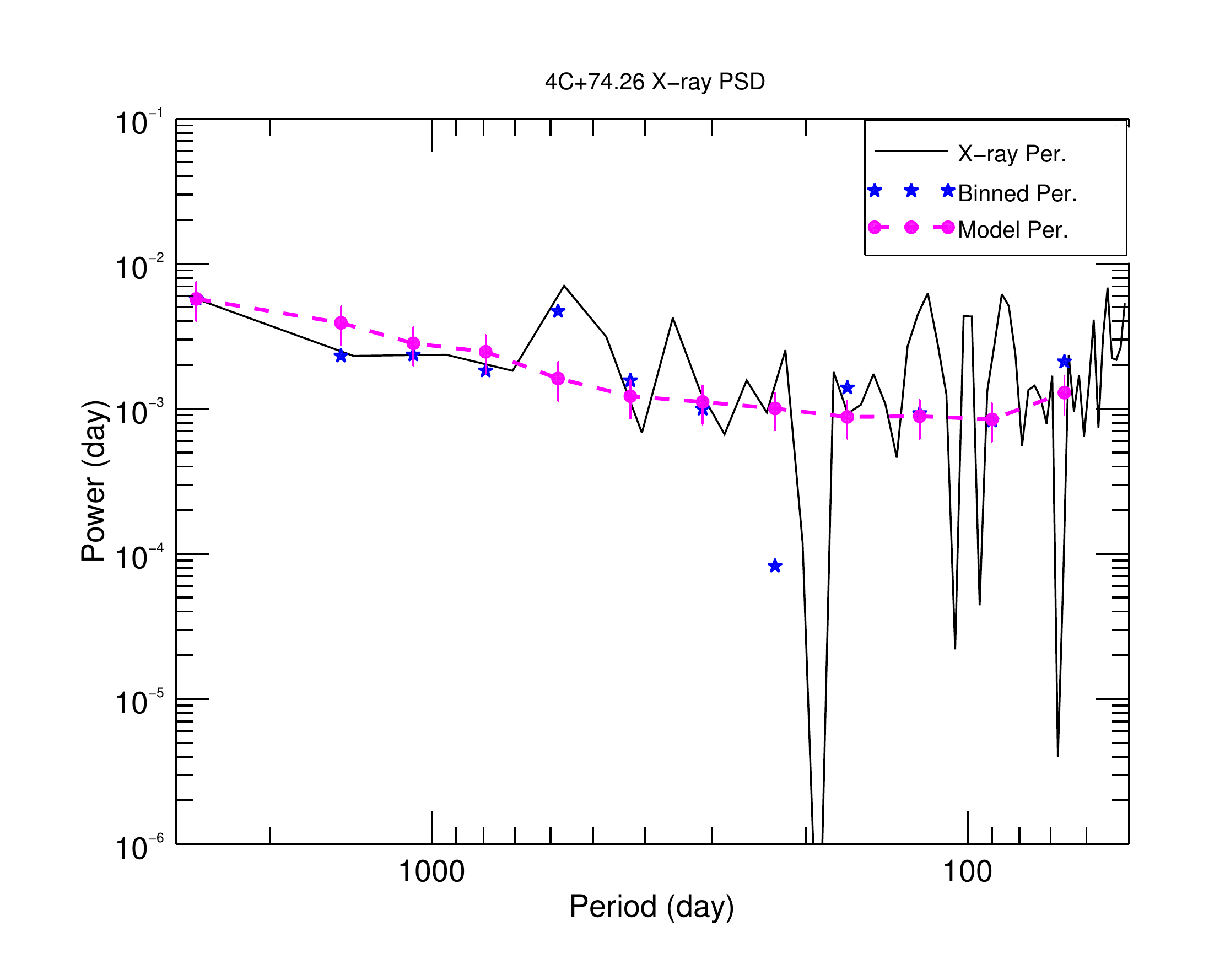}
\caption{The unbinned (black curve) and binned (blue stars) periodograms for the optical, radio and hard X-ray light curves of 4C+74.26 are shown in the top, middle, and bottom panels, respectively. The corresponding magenta curves represent the reference models with the spectral slopes $\beta =1.6\pm0.2$ (optical), $1.4\pm0.1$ (radio) and $0.9\pm0.2$ (X-ray), along with the associated errors estimated as $1 \sigma$ deviations of the distribution in the simulated periodograms at a given frequency.}
\label{power-law}
\end{figure}

\begin{figure*}[th!]
\begin{center}
\begin{tabular}{cc}
\includegraphics[angle=0,width =\columnwidth]{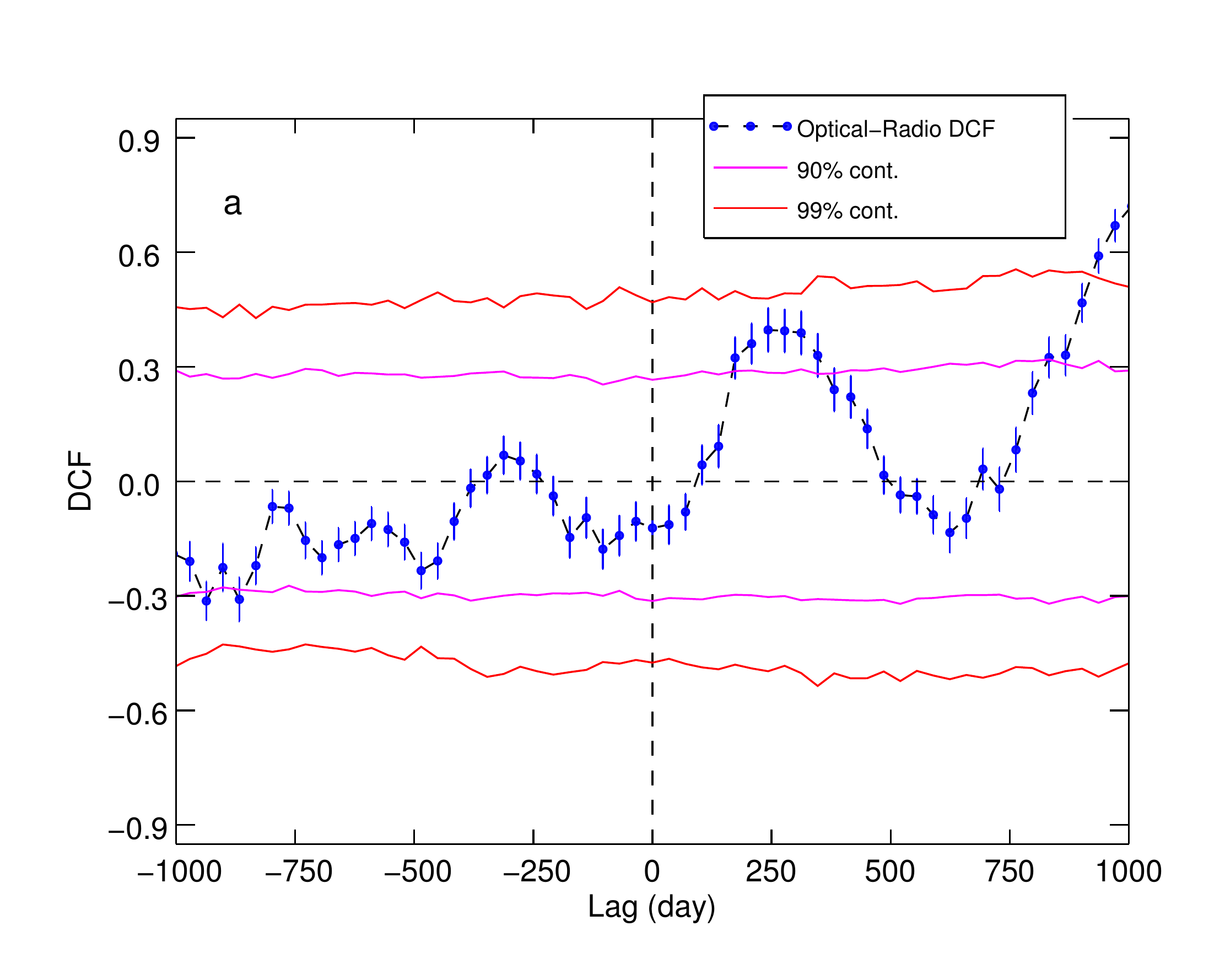}&
\includegraphics[angle=0,width =\columnwidth]{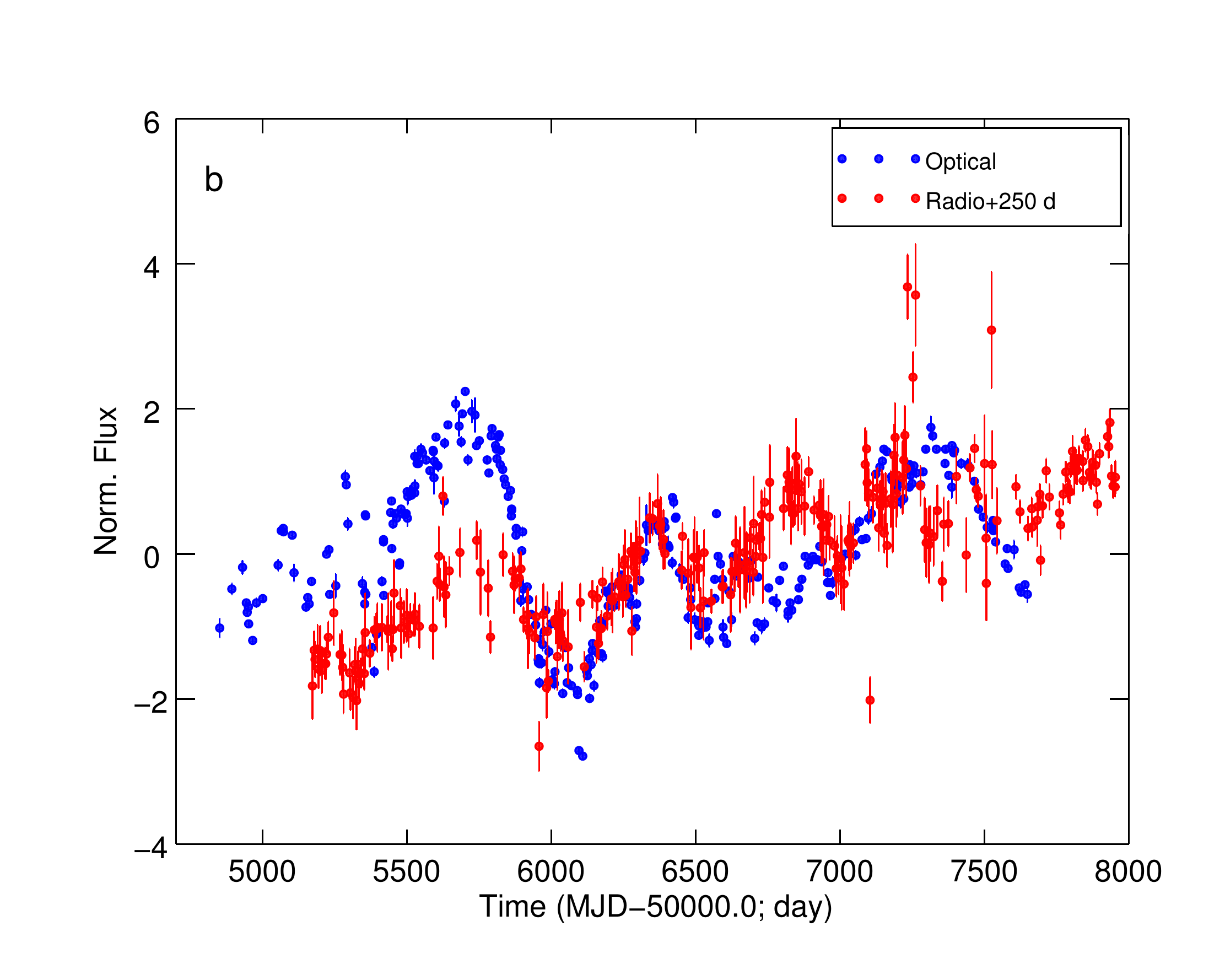}\\

\vspace{0.3cm}

\includegraphics[angle=0,width =\columnwidth]{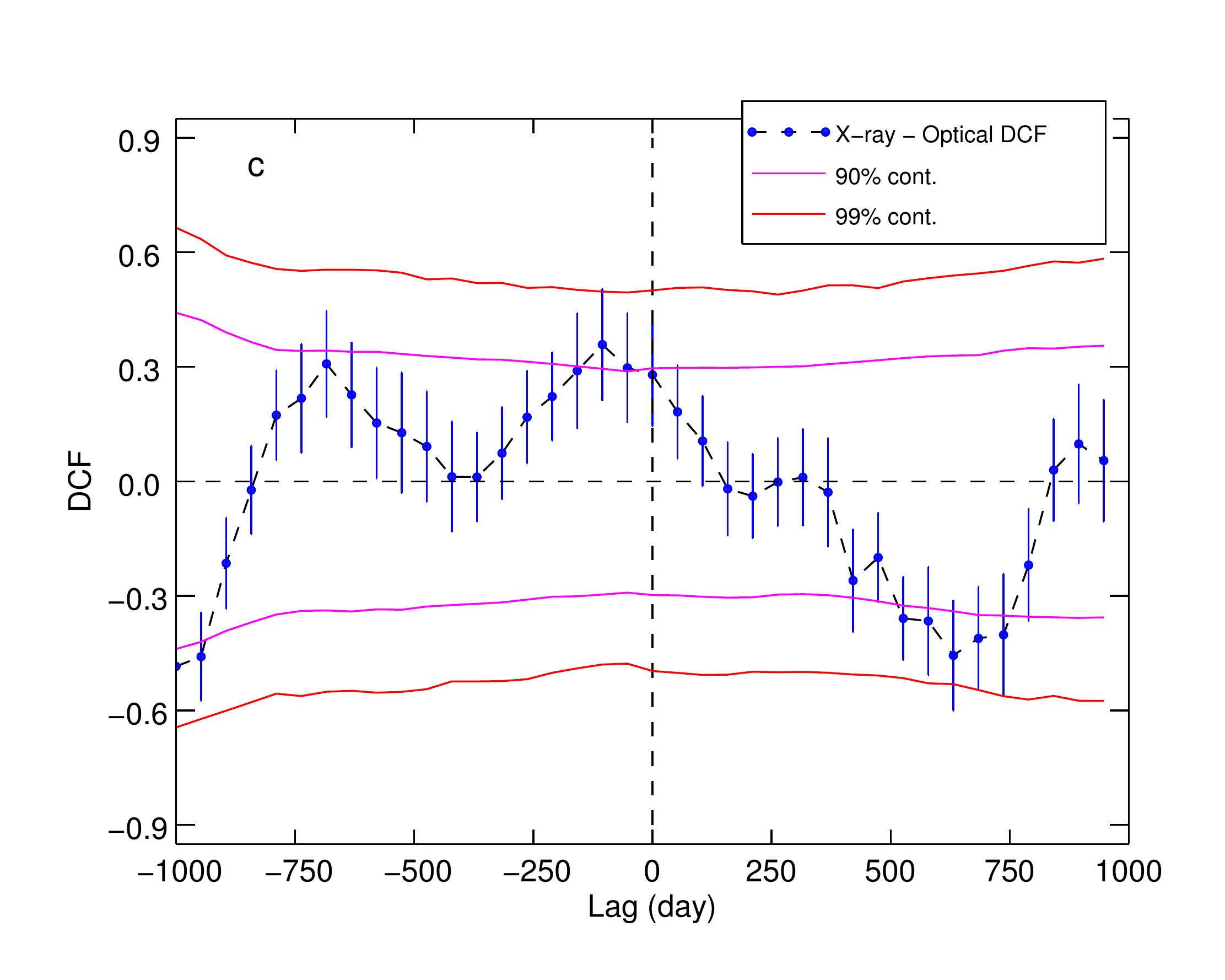}&
\includegraphics[angle=0,width =\columnwidth]{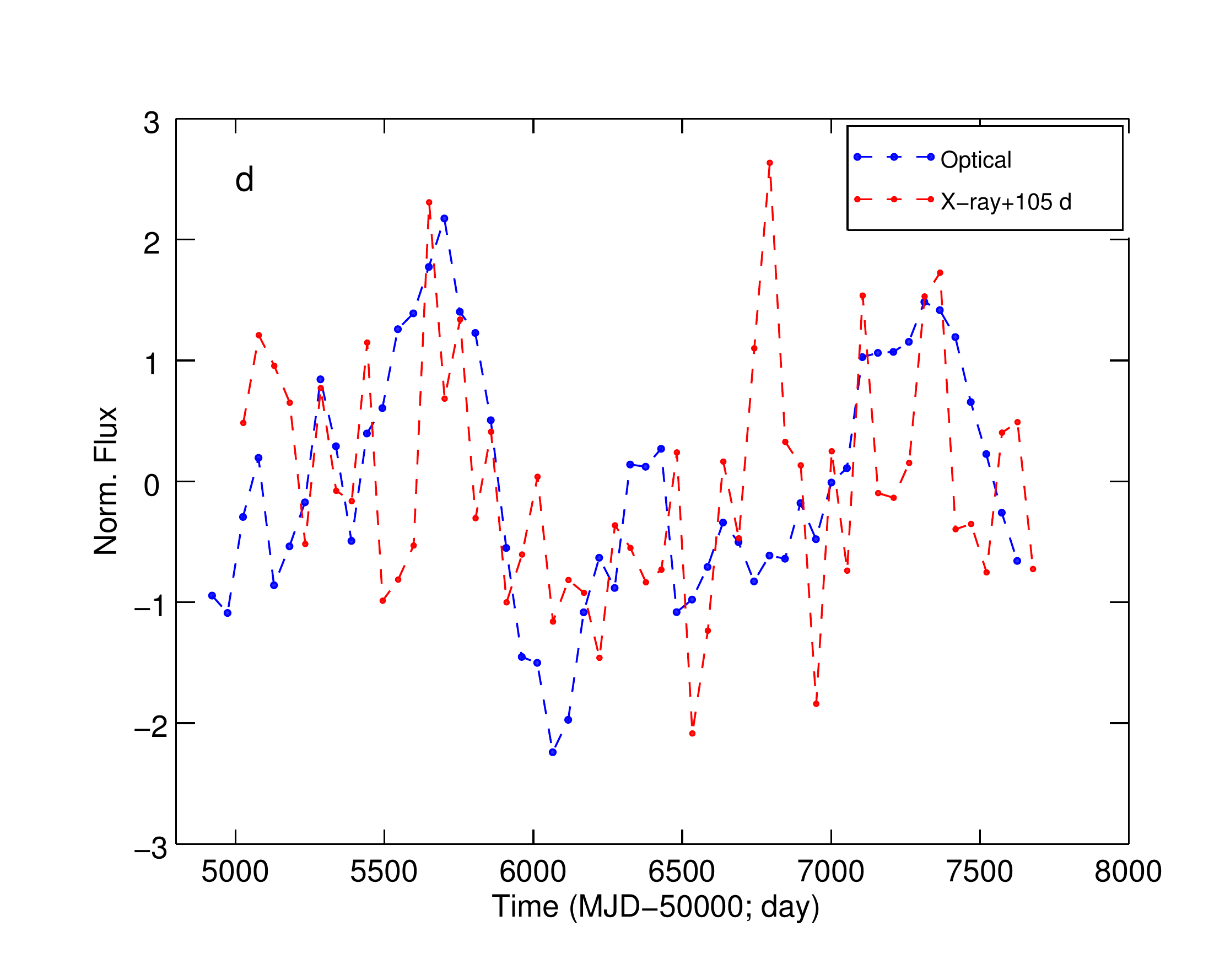}\\

\vspace{0.3cm}

\includegraphics[angle=0,width =\columnwidth]{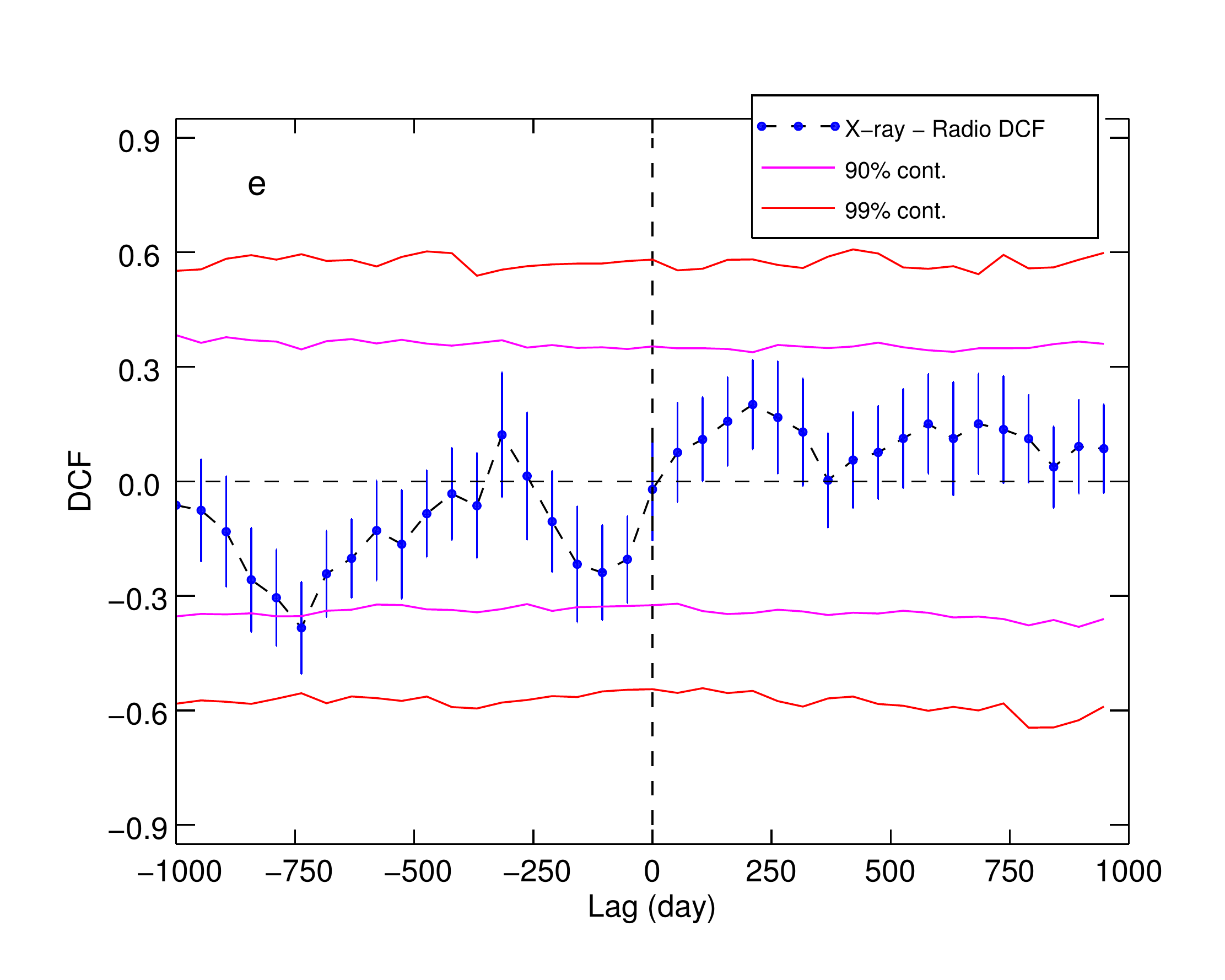}&
\includegraphics[angle=0,width =\columnwidth]{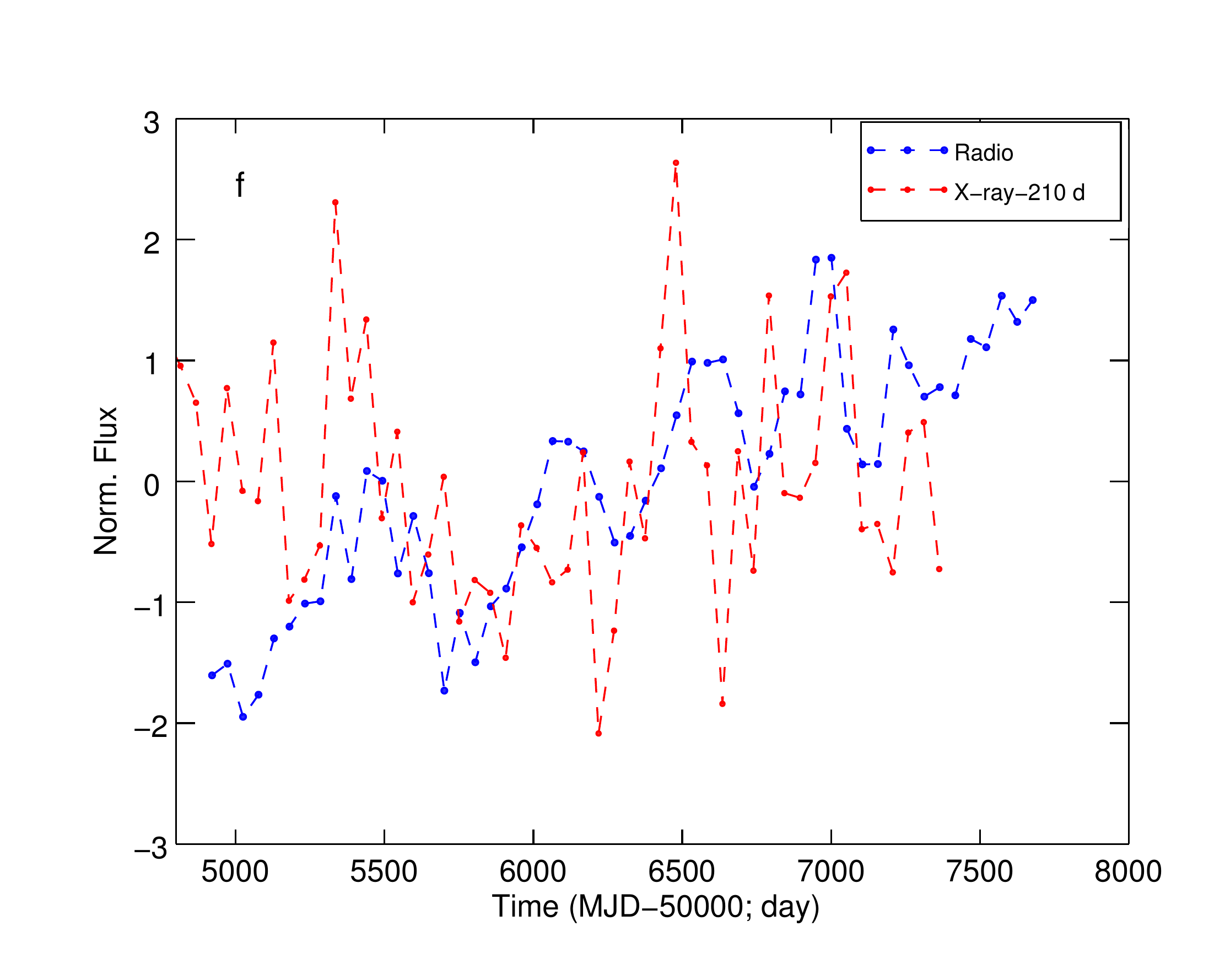}\\
\end{tabular}
\caption{Multifrequency correlations of 4C\,+74.26. In the left panels, the DCFs between the optical and the radio (panel `a'), the X-ray and the optical (panel `c'), and the X-ray and the radio (panel `e') light curves are shown with blue symbols. Here positive lag indicate the hard lag. The magenta and red curves in the left panels denote the 90\% and the 99\% local significance levels, respectively, calculated from the distributions of the DCFs between the simulated red-noise light curves (see Section~\ref{sec:PSD}). Right panels show the corresponding normalized light curves shifted by the most prominent time lags emerging from the DCF analysis, and in particular the radio lightcurve shifted by +250\,d (panel `b'), the X-ray lightcurve by +105\,d (panel `d'), and finally the X-ray lightcurve by --210\,d (panel `f').}
\label{CCF}
\end{center}
\end{figure*}

\section{Analysis}
\label{sec:analysis}
 
\subsection{Power Spectral Density \label{sec:significance} }
\label{sec:PSD}
 
In order to test the significance of the correlations between the optical, radio, and X-ray flux changes in 4C\,+74.26, using a robust statistical tool involving Monte Carlo simulations, first of all the variability of the target has to be quantified with the Power Spectral Density (PSD). It should be pointed out that, in general, AGN continuum variability at various frequencies has been established -- using various techniques and methods -- to be of a colored noise type \citep[see in this context, e.g.,][]{Markowitz2003,Chatterjee08,Kelly09,Kelly11,Kastendieck2011,Edelson2013,Sobolewska14,bhatta16b,Bhatta2017,Abdalla17,Ahnen17,Goyal17,Goyal18}. Here, for the given source light curves, we therefore modelled the PSDs as single power-laws plus constant components corresponding to the Poisson noise levels, $P_{f} = A \, f^{-\beta}+const$, where $f$ denotes the temporal frequency. The details of the Discrete Fourier Periodogram modeling, involving extensive simulations to estimate the best-fit power-law model following the method described in \citet{Uttley2002}, can be found in \citet{bhatta16a,bhatta16b}. Figure\,\ref{power-law} presents the periodograms and the corresponding PSD best-fit models with spectral slopes $\beta=1.6\pm0.2, 1.4\pm
0.1$ and $0.9\pm 0.2$ for the optical, radio and X-ray PSDs, respectively.

\subsection{Discrete Correlation Function}
\label{sec:DCF}
 
The discrete cross-correlation function (DCF) is one of the widely used methods to estimate cross correlations between unevenly sampled light curves of astrophysical sources (\citealt{Edelson1988}; see also \citealt{Bhatta2018a}). In the case of 4C\,+74.26, we evaluated the DCFs between the optical and the radio, the X-ray and the optical, and the X-ray and the radio flux changes. The left panels (a, c, and e) in Figure~\ref{CCF} present the resulting DCFs, while the right panels (b, d, and f) show the corresponding normalized light curves shifted by the most prominent time lag, for illustration. The significance of the DCF peaks was estimated using a large number of simulated light curves based on the PSD analysis discussed in Section~\ref{sec:PSD}. In particular, for the best-fit PSD parameters, 10,000 simulated light curves were generated in each frequency band (i.e., X-ray, optical, and radio bands) and sampled according to the corresponding observations.  Subsequently DCF analysis was performed between the pairs of the simulated light curves. From the resulting distribution of the DCFs, local 90\% and  99\% significance contours, denoted respectively by the magenta and the red curves in the figure (left panels), were estimated \citep[see][]{Max-Moerbeck2014b,Aleksic2015}. 

The uncertainty in the observed lags were estimated using the flux randomization subset selection (FR/RSS) method  (see \citealt{Peterson1998}, and also \citealt{Bell2011}): from the original light curves, 10,000 random subsets containing only  63\% of the total observations were selected, and the distributions of the peak DCFs were obtained; the standard deviation of this distribution of the lags corresponding to the peak DCFs was adopted as the measure of the lag uncertainty.

As shown in Figure~\ref{CCF} and summarized in Table~\ref{tab:DCF}, the optical and the radio light curves of 4C\,+74.26 are correlated with the time lag of $\sim 250\pm42$\,d at the 96\% significance level, even though the largest significant DCF value is only $\sim 0.4$. In addition to the aforementioned local significance, we also evaluated a \emph{global significance} of the observed lag to be 98\%. Global significance accounts for the fact that, while searching the correlations over a wide range of time lags, we do not have a priori knowledge of what significance lag would emerge out from the analysis (\citealt{Bell2011},  see also \citealt[][]{Bhatta2017}).  Using the distribution of the simulated DCFs between optical and the radio frequency, global significance was estimated as a fraction of the DCFs  below the observed local significance level i.e. the 96\% level, at all the lags considered. It is therefore very likely that here we do see, for the first time in any AGN, a direct relation between the disk and the jet emission components, although the fact that radio (jet) leads the optical (disk) may seem counter-intuitive at first. We address this point in more detail in Section~\ref{sec:discussion} below.

\begin{figure}[!t]
\includegraphics[width=0.72\columnwidth, angle=-90]{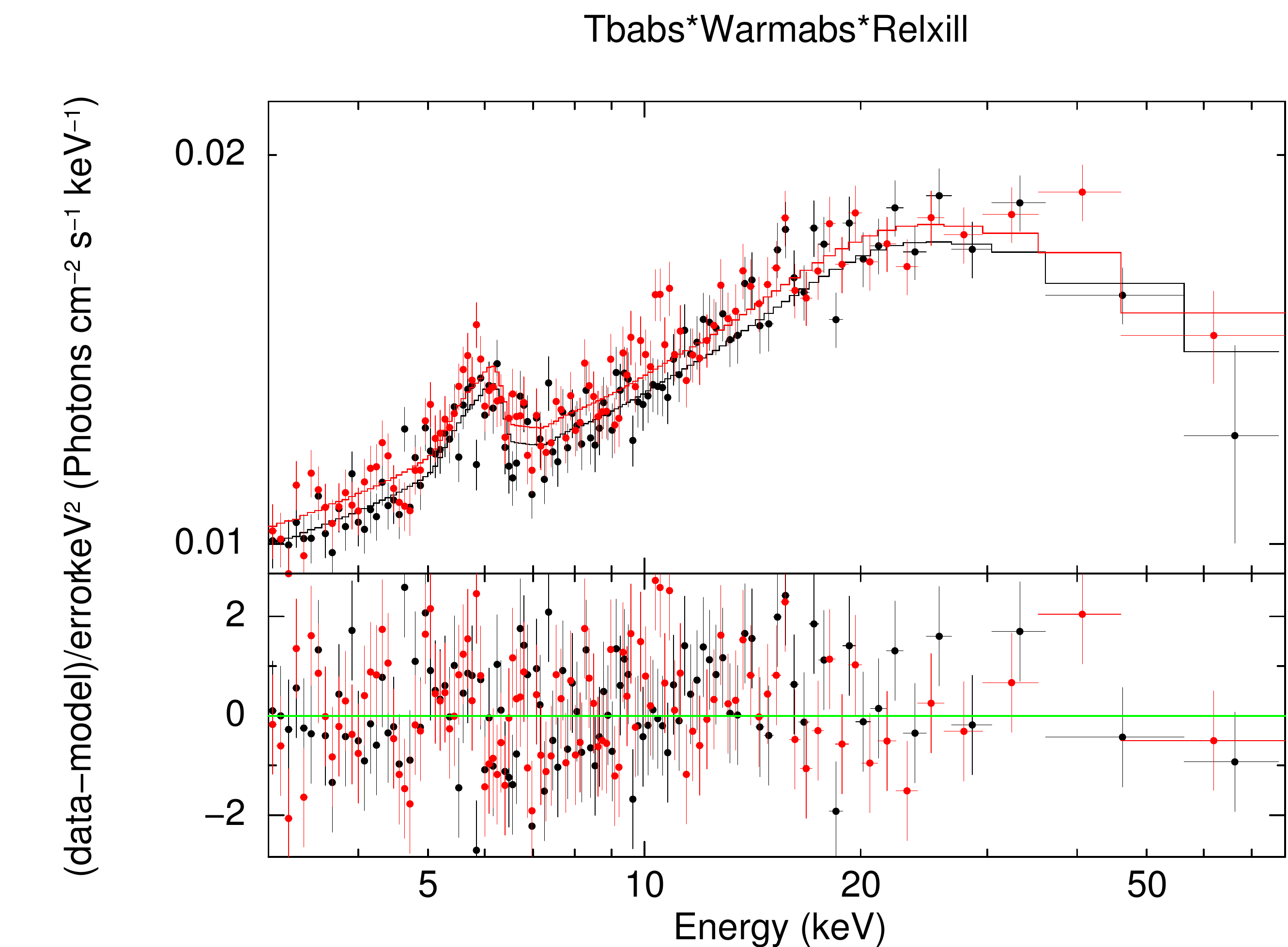}\\
\caption{Spectral modeling of the \nustar data for 4C\,+74.26 (obs. ID 60001080006, 2014-10-30), assuming the convolved model \texttt{tbabs*warmabs*relxill} (upper panel), with the corresponding residuals (lower panel). For the model description and the best-fit model parameters see Section~\ref{sec:Xrays}. Red and black symbols/curves correspond to the \nustar modules FPMA and FPMB.}
\label{fig:nustar}
\end{figure}

It should be pointed out that, as seen in the Figure 4a, the correlation coefficient between the two light curves appears to be the largest near +1,000\,d. However, such seemingly strong correlations appearing in the DCF diagrams at the lags larger than $\sim$\,half the duration of a given light curve, should be treated with extreme caution in the case of a coloured-noise type of a source variability, as for such the largest-amplitude flux changes expected on the longest timescales can introduce false correlation ``trends'' \citep[e.g.,][]{Press1978,Max-Moerbeck2014a}. And in general, for larger and larger lags the number of pairs of the light curve data points become smaller and smaller, and this may also lead to spurious peaks in the DCF diagrams.

Unfortunately, the variability seen in hard X-ray band by {\it Swift}-BAT turns out heavily dominated by the Poisson noise on short (daily/weekly) timescales, and therefore we had to use the 100\,d bins when correlating it with the optical and the radio light curves. With such, the resulting correlation time lags are detected at the local significance levels of only $<90\%$.

\subsection{X-ray Spectral Analysis}
\label{sec:Xrays}

An in-depth analysis of the rich X-ray dataset for 4C\,+74.26, using the most recent observations with {\it XMM-Newton}, {\it Chandra}, and \nustar, has been reported by \citet{DiGesu2016} and \citet{Lohfink2017}. Here we re-analyze the \nustar data for the target in the hard X-ray regime, focusing predominantly on the characterization of the main parameters of the underlying accretion disk, in particular its inner radius, in the context of the multi-wavelength correlation studies reported in the previous Section~\ref{sec:DCF}.

4C\,+74.26 has been observed with \nustar four times between September 2014 and December 2014. Because of the source variability seen in the {\it Swift}-BAT light curve, for our spectral analysis we have selected only the longest \nustar exposure of 90.925\,ks (obs. ID 60001080006, 2014-10-30), and fitted simultaneously the FPMA and FPMB spectra with the \texttt{XSPEC} \citep{Arnaud1996}. We used the latest update of the \texttt{RELXILL} [v1.0.2] model\footnote{\url{http://www.sternwarte.uni-erlangen.de/~dauser/research/relxill/}}, which convolves the \texttt{XILLVER} reflection code with the \texttt{RELLINE} code \citep{Dauser2010, Dauser2013}, providing as such an accurate prescription for a reflection from a Compton-thick disk illuminated by a coronal power-law continuum, with a good treatment of strong gravity effects near the event horizon of a supermassive black hole \citep[for details see][]{Garcia2010,Garcia2013}. In addition, in order to account for the presence of a warm absorber in the source \citep[e.g.][]{DiGesu2016},  as well as the Galactic absorption, we combined \texttt{RELXILL} with the \texttt{TBABS} \citep{Wilms2000} and \texttt{WARMABS} models.  
 
Guided by the previous studies reported in the literature, in the modelling we fixed the Galactic column density at  $N_{\rm H,\,Gal} = 2.31 \times 10^{21}$\,cm$^{-2}$ \citep[see][]{DiGesu2016}. For the single-zone warm absorber with the outflow velocity of 3600\,km\,s$^{-1}$, we assumed the column density $N_{\rm H,\,wa} = 3.5 \times 10^{21}$\,cm$^{-2}$, the ionization parameter $\log \xi_{\rm wa} = 2.6$, and the turbulent velocity $v_{\rm turb}=100$\,km\,s$^{-1}$ \citep{DiGesu2016}. For the coronal component and the accretion disk itself in the \texttt{XILLVER} model, we assumed the emissivity indices $Index_1=Index_2=3$, the outer disk radius $R_{\rm out}/R_{\rm g} = 10^3$ (where $R_{\rm g}$ is the gravitational radius), the inclination $i=45\deg$, and the solar iron abundance $A_{\rm Fe}=1$; following \citet{Lohfink2017}, we also fixed the cut-off energy as $E_{\rm cut} = 180$\,keV, and considered only the maximum spin of the supermassive black hole, namely $a=0.998$. The remaining model free parameters are the inner disk radius $R_{\rm in}$, the power-law index of the incident spectrum $\Gamma$, the ionization of the accretion disk $\xi$, the reflection fraction $R_f$, and the normalization.
 
\begin{figure}[t!]
\includegraphics[width=0.77\columnwidth, angle=-90]{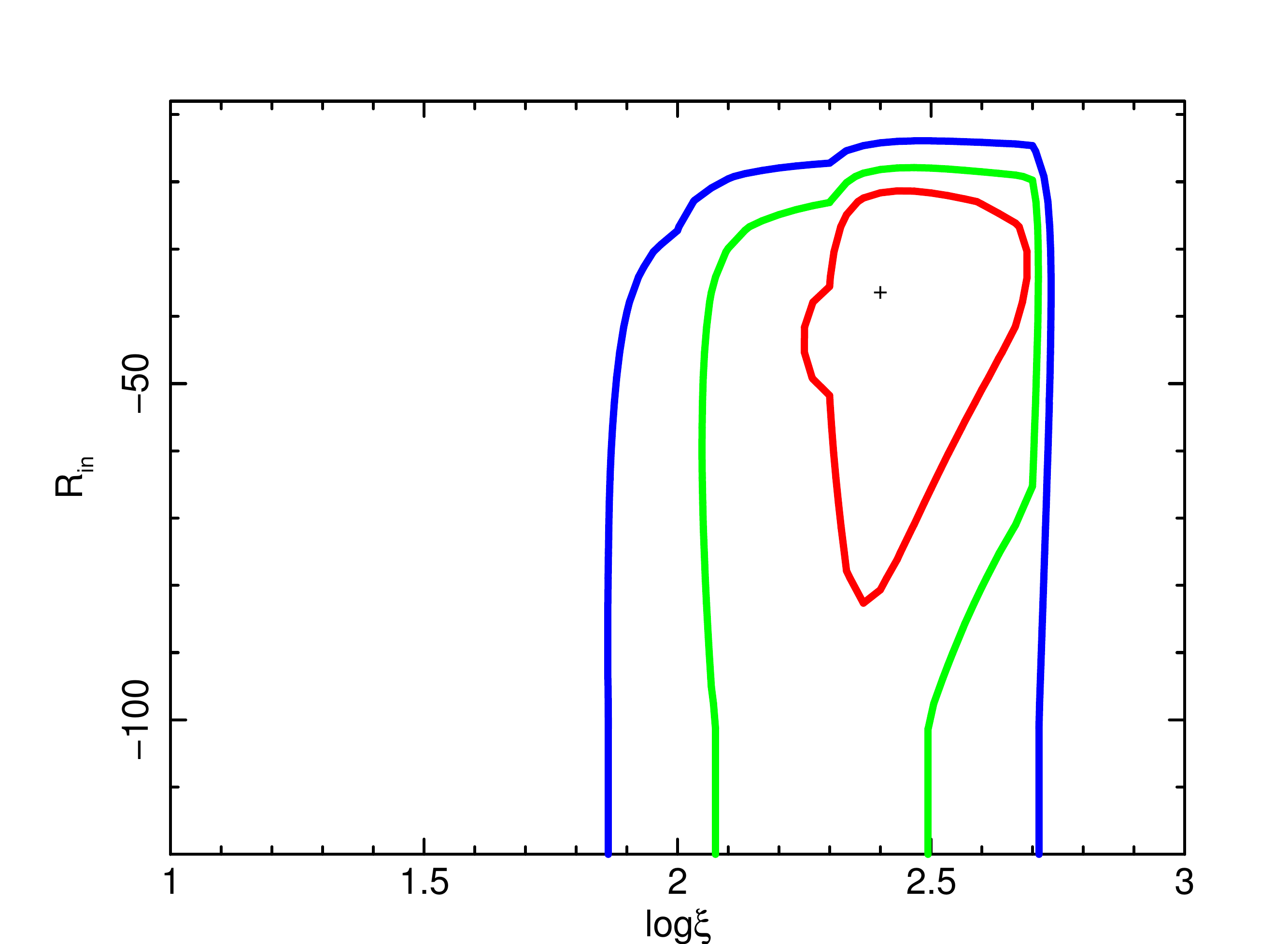}\\
\vspace{0.7cm}
\includegraphics[width=0.77\columnwidth, angle=-90]{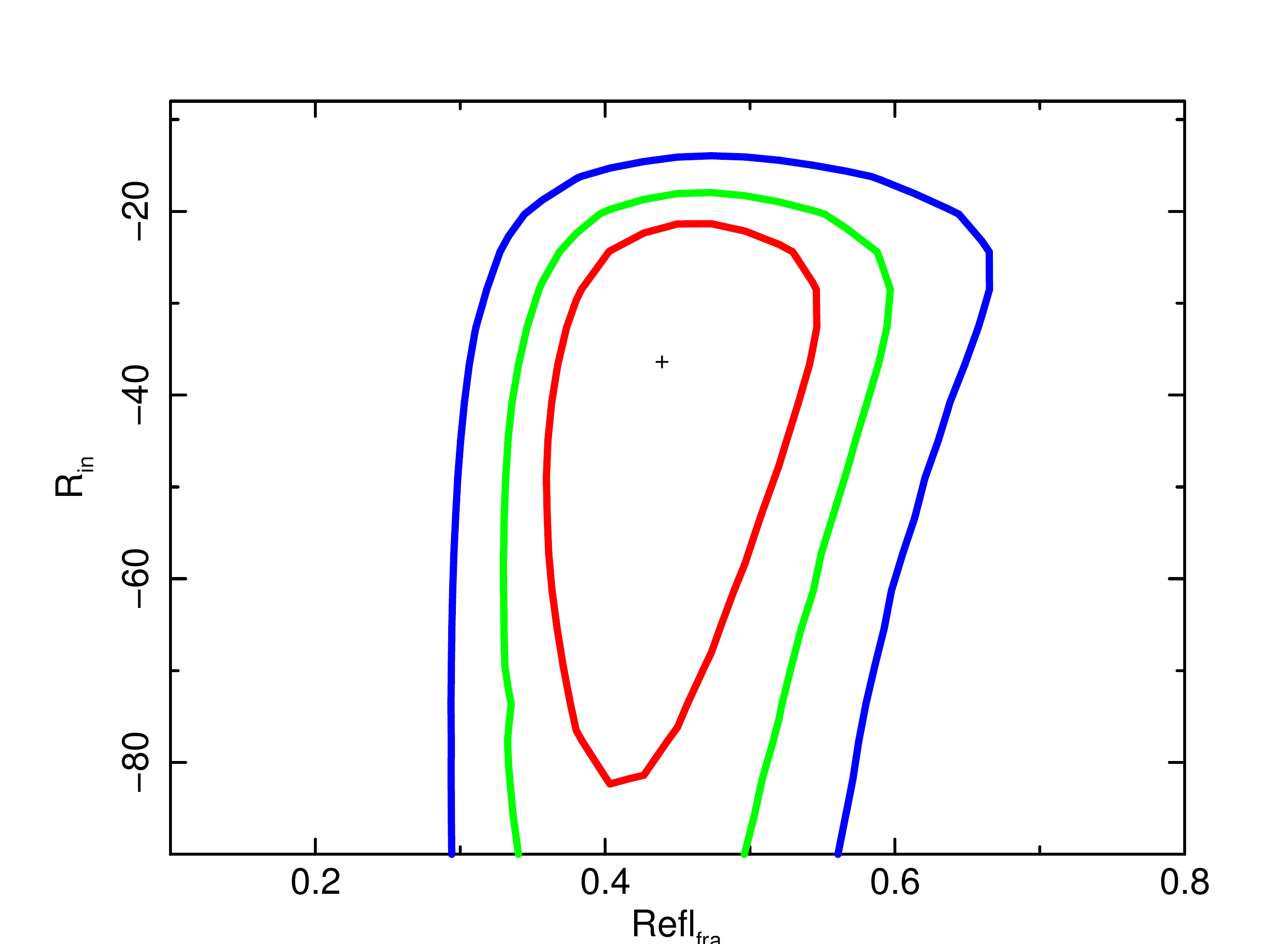}\\
\vspace{0.7cm}
\includegraphics[width=0.77\columnwidth, angle=-90]{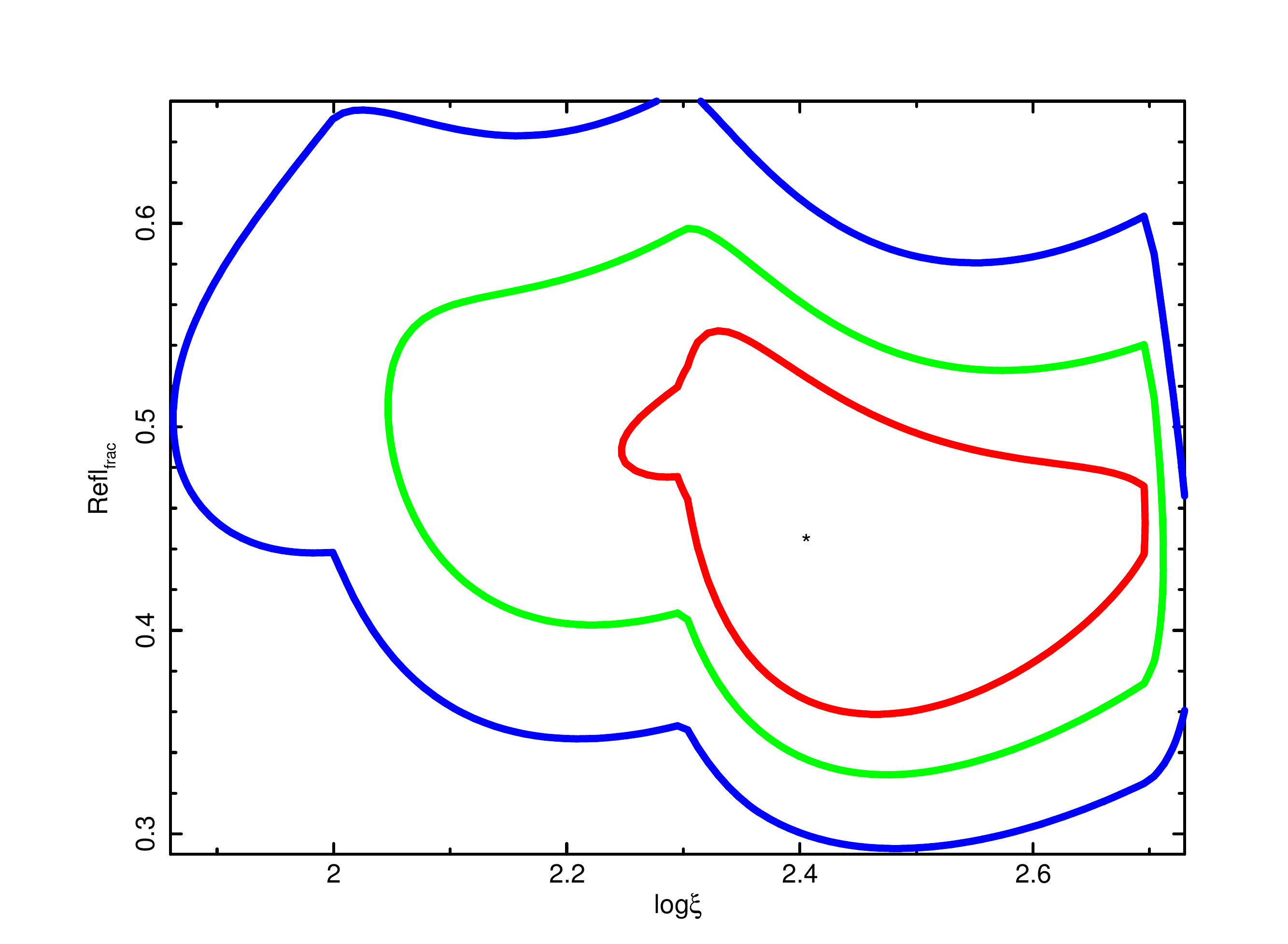}\\
\caption{Confidence contours (see Section~\ref{sec:Xrays}) for $\log \xi-R_{\rm in}$ (upper panel), $R_f -R_{\rm in}$ (middle panel), and $\log \xi-R_{\rm f}$ (lower panel); red, green, and blue curves represent 1, 2, and $3 \sigma$ contours respectively. The negative sign seen in the figure indicates that $R_{\rm in}$ is expressed in the units of $R_{\rm ISCO}$, according to the nomenclature adopted in the \texttt{XILLVER} model.}
\label{fig:contours}
\end{figure}

Our best-fit model ($\chi^{2}/{\rm dof}=0.99$), presented in Figure~\ref{fig:nustar}, returns $R_{\rm in}/R_{\rm ISCO} = 35 \pm 22$, where $R_{\rm ISCO}$ stands for the innermost stable circular orbit, $\Gamma = 1.83\pm 0.02$, $\log \xi  = 2.47 \pm 0.17$, and $R_{\rm f}=0.49\pm0.07$. In order to estimate even more robustly the model free parameters of our main interest, we studied the $\log \xi-R_{\rm in}$, $R_f -R_{\rm in}$, and $\log \xi-R_{\rm f}$ confidence contours (Figure~\ref{fig:contours}), obtaining the $3\sigma$ bounds $R_{\rm in}/R_{\rm ISCO}=35^{+40}_{-16}$, $\log \xi = 2.47^{ +0.17}_{-0.19}$, and $R_{\rm f}=0.44^{+1.3}_{-0.09}$. This implies that the accretion disk in 4C\,+74.26 is indeed most likely truncated at the radius of about a few tens of the ISCO; note that for the maximum black hole spin assumed here, $R_{\rm ISCO} \simeq R_{\rm g}$.

    
\section{Discussion and Conclusion}
\label{sec:discussion}

In the current work, with an aim to explore the disk-jet connection in AGN, we performed a multi-frequency study of 4C+74.26 using multifarious methods, including the broad-band spectral energy distribution modelling, the variability power spectral density analysis, the discrete cross-correlation of the multi-wavelength light curves of the source, and the detailed X-ray spectroscopy enabled by the high-quality and broad-band \nustar data.

A comparison of the source broad-band SED with those of other broad-line radio galaxies and radio-loud quasars compiled and analyzed by \citet{Kataoka2011}, indicates that, in the case of 4C\,+74.26, the observed infrared--to--X-ray continuum could not be accounted by a non-thermal jet emission, but instead has to be accretion-related, including the direct disk emission component (optical frequencies), plus the disk emission re-processed by the circumnuclear dusty torus (infrared) and the disk corona (X-rays). This is illustrated in Figure~\ref{SED} by matching the SED datapoints with the normalized ``accretion-related'' quasar template from \citet[red long-dashed curve]{Koratkar99}, and the ``jet-related'' 3C\,273 template from \citet[blue short-dashed curve]{Soldi08}.

For a comparison with our PSD results, we note that PSDs of nearby radio-quiet Seyferts have been investigated extensively at X-ray ($1-10$\,keV) frequencies during the last two decades. In the frequency range $10^{-8}-10^{-3}$\,Hz (timescales from years to hours), these PSDs are typically modelled by a broken power-law model $P(f) \propto f^{-\beta_\textrm{L}}$ or $f^{-\beta_\textrm{H}}$ below or above a break frequency $f_0$, respectively; $f_0$ typically corresponds to timescales of hours--days, $\beta_\textrm {L}$ is typically $\leq 1$, and $\beta_\textrm {H}$ is typically $\sim2-3$, with the timing properties showing strong similarities to those of black hole X-ray Binaries \citep[see, e.g.,][]{Uttley2002,Markowitz2003}. These include the X-ray PSDs of several radio-loud but non-blazar AGN, in particular 3C\,390.3 and 3C\,120 \citep{Gliozzi09,Marshall09}.\footnote{For a summary of  multi-wavelength PSD analysis for blazar sources, see \citet{Goyal17,Goyal18} and references therein.} The \textit{Swift}/BAT (10--150\,keV) lightcurves of Seyfert galaxies analyzed by \citet{Shimizu13} also yielded flat PSD power-law slopes, $\beta_L \lesssim 1$, at timescales longer than days/weeks. Regarding optical-band AGN PSDs, \emph{Kepler} monitoring of several radio quiet AGN has revealed steep power-law slopes $\sim 3$ over the range $10^{-7} - 10^{-5}$\,Hz \citep{Mushotzky11}. On the other hand, optical PSDs of several radio-loud AGN also based on \emph{Kepler} data are consistent with somewhat flatter power-law forms with indices $\lesssim 2.0$ \citep{Wehrle13,Revalski14}.

A cross-correlation study between the long-term optical (R-band) and the higher-frequency radio (15\,GHz) light curves of the broad-line radio galaxy/quasar 4C\,+74.26, indicates that the radio (jet) emission leads the optical (disk) emission by $\sim 200-300$\,days. This seems counter-intuitive at first, since one could expect that any changes in the accretion disk should come first, and their imprint on the relativistic outflow follows only later. And since relativistic jets in AGN are widely believed to be launched by magnetic fields supported by the disks and entering the ergospheres of spinning black holes \citep{Blandford77}, such changes could be naturally identified with magnetic perturbations within the innermost parts of the accretion disk. However, it takes some time for the accretion disk itself to respond radiatively to magnetic fluctuations, and for a standard ``geometrically thin/optically thick'' disk model \citep[e.g.][]{Shakura1973}, the characteristic timescale for such a response is the thermal timescale, $\tau_{\rm th} \sim \alpha^{-1} \, \tau_{\rm dyn}$, where $\alpha$ is the disk viscosity, and $\tau_{\rm dyn} = r^{3/2} / \sqrt{G \mathcal{M}_{\rm BH}}$ is the orbital time scale at the distance $r$ from the central black hole with mass $\mathcal{M}_{\rm BH}$ {\bf \citep[see, e.g.,][]{Czerny06}}. In the particular case of 4C\,+74.26, with a black hole mass of $\mathcal{M}_{\rm BH} \simeq 10^{9.6}M_{\odot}$ \citep{Gofford2015}, meaning the gravitational radius $R_{\rm g} \sim 6 \times 10^{14}$\,cm, and assuming the disk viscosity $\alpha \sim 0.1$, this thermal timescale around the inner disk radius $R_{\rm in} \sim 35\,R_{\rm g}$ emerging from the \nustar data modelling, reads as
\begin{equation}
\tau_{\rm th} \sim \alpha^{-1}  \, \frac{R_{g}}{c}  \left(\frac {R_{\rm in}}{R_{g}}  \right)^{3/2} \sim 470\,{\rm d} \, .
\end{equation}

Meanwhile, the timescale for a magnetic perturbation to travel the projected distance $\ell_{\rm proj}$ along the jet with the relativistic bulk velocity $\beta_j \sim 1$, is
\begin{equation}
\tau_{\rm j} = \frac{\ell_{\rm proj}}{c \beta_{\rm j}} \, \frac{1 - \beta_{\rm j}  \, \cos \theta} {\sin \theta} \, ,
\end{equation}
where $\theta$ is the jet viewing angle {\bf \citep[e.g.,][]{Stawarz04}}. For example, with the anticipated $\theta=45\deg$ and the jet bulk Lorentz factor $\Gamma_j \simeq 10$, this timescale reads as $\tau_{\rm j} \lesssim 50$\,d for the projected distance $\ell_{\rm proj} \lesssim 0.1$\,pc, i.e. where one should expect to find the dominant production zone of the 15\,GHz photons in the 4C\,+74.26 jet. This supports the idea that both the observed disk variability at optical wavelengths and the jet variability at cm wavelengths are generated by magnetic field fluctuations within the innermost parts of the disk, and that the lag we observe between the disk and the jet emission is related to a much delayed radiative ``response'' of the disk to magnetic field fluctuations when compared with relativistic jet timescales at sub-parsec distances from the core, $\tau_{\rm th} - \tau_{\rm j} \sim \mathcal{O}(100\,{\rm d})$. 

We note in this context that numerical (MHD shearing box) simulations by \citet{Hirose09} revealed that indeed ``magnetic fluctuations lead radiation energy fluctuations by 5--15 orbits, a little less than a thermal time'', and that such a delay is also consistent with the stochastic modelling of optical variability power spectra in quasar sources by \citet{Kelly09,Kelly11}. 

It should be also noted at this point, that at the source redshift, the observed R-band photons correspond to the blackbody temperature of 4,900\,K. Taking the well-known formula for the temperature of a standard accretion disk {\bf \citep[e.g.,][]{Lasota16}},
\begin{equation}
\left(\frac{T}{10^{5.8}\,{\rm K}}\right) \simeq \left(\frac{\dot{M}}{\dot{M}_{\rm Edd}}\right)^{1/4} \!\! \left(\frac{\mathcal{M}_{\rm BH}}{10^8 M_{\odot}}\right)^{-1/4} \!\! \left(\frac{r}{2 R_{\rm g}}\right)^{-3/4} \, ,
\end{equation}
where $\dot{M}/\dot{M}_{\rm Edd}$ is the accretion rate in the Eddington units (see below), one may therefore find that, for the source parameters as discussed above, the radius corresponding to the temperature of 4,900\,K is $r \sim 200 \, R_{\rm g}$, which is larger than the inner disk radius $R_{\rm in}$ estimated above by means of the spectral modelling of the X-ray data. However, not all of the emission of the accretion disk observed at optical frequencies is produced exclusively at large disk radii. In fact, one may easily check that, in the observed R-band, the segment of the disk extending from the inner disk radius $R_{\rm in} \sim 35 \, R_{\rm g}$ up to $\sim 50 \, R_{\rm g}$ contributes as much as $\sim 10\%$ of the total disk intensity in the source,
\begin{equation}
I_{\nu} \propto \int_{R_{\rm in}}^{R_{\rm out}} \!\!\! dr \, r \, B_{\nu}\!(T) \propto R_{\rm in}^2  T_{\rm disk}^{8/3} \nu^{1/3} \int_{y_{\rm min}}^{y_{\rm max}} \frac{y^{5/3} \, dy}{\left(e^y-1\right)}
\end{equation}
where $B_{\nu}\!(T)$ denotes the intensity of a blackbody, $T_{\rm disk}= 4.8 \times 10^5 \, (R_{\rm in}/R_{\rm g})^{-3/4}$\,K, $y_{\rm min}=h\nu/kT_{\rm disk}$, and $y_{\rm max}=y_{\rm min} (R_{\rm out}/R_{\rm in})^{3/4}$.

 Still, one has to keep in mind that the above interpretation is not unique, that the significance of the optical/radio lags revealed by our cross-correlation analysis is still only modest, and also that the accretion disk parameters estimated in support of our interpretation via the X-ray modelling are subjected to relatively large uncertainties, despite an excellent quality of the \nustar data and state-of-the-art spectral models applied. For example, with the $3\sigma$ bounds $R_{\rm in}/R_{\rm ISCO}=35^{+40}_{-16}$ obtained in Section\,\ref{sec:Xrays}, one \emph{formally} obtains a rather wide range $190$\,d\,$\leq \tau_{\rm th} \leq 15000$\,d. Also, in the framework of the scenario proposed, one could expect a significant correlation between the hard X-ray and optical and/or radio light curves. The fact that no such correlations are seen above 90\% confidence levels (see Table\,\ref{tab:DCF}) seems, on the other hand, to be predominantly due to a rather low signal-to-noise ratio in the {\it Swift}-BAT data (see Section\,\ref{sec:DCF}); at the same time, we note that any correlated X-ray variability could in principle be diluted by an additional torus-reflected component (not considered in our \nustar data modelling).

Let us finally comment here on the overall energetics of 4C\,+74.26. The $14-195$\,keV flux of the source gives a value of the hard X-ray luminosity $L_{\rm X} \simeq 1.3 \times 10^{45}$\,erg\,s$^{-1}$, and hence, following \citet{Ichikawa2017}, the accretion-related bolometric luminosity of $L_{\rm acc} \simeq 8.6 \times 10^{46}$\,erg\,s$^{-1}$, meaning that the accretion rate in Eddington units is $\dot{M}/\dot{M}_{\rm Edd} \sim 0.2$. This is roughly consistent with the 3.4\,$\mu$m--22\,$\mu$m (WISE) luminosity of 4C\,+74.26, namely $L_{\rm IR} \simeq 5.3 \times 10^{45}$\,erg\,s$^{-1}$, for the anticipated $\sim 1\%-10\%$ of the disk emission being reprocessed by the hot dusty torus. Meanwhile, the jet kinetic power in 4C\,+74.26 turns out to be relatively low. This jet kinetic power can be estimated the most robustly by means of a spectral modelling of the radio emission of giant lobes in the source, accounting for the complete history of the energy evolution of relativistic particles injected into the expanding lobes by a pair of relativistic jets \citep[see][and the discussion therein]{Machalski07}. Using the data gathered in \citet{Machalski2011}, and the evolutionary DYNAGE code by \citet{Machalski07}, we have performed such a modelling, assuming the jet inclination of $45\deg$. This resulted in an estimate for the age of the radio source of $\simeq 87$\,Myr, and the total (time-averaged) jet kinetic power of $L_{\rm j} \simeq 4 \times 10^{44}$\,erg\,s$^{-1}$, implying a rather low jet production efficiency in the source, $L_{\rm j} / L_{\rm acc} \sim 0.01$. Interestingly, the total kinetic luminosity of the ionized nuclear outflow in 4C\,+74.26 has been found to be even lower than this, namely $L_{\rm out} \sim 1.5  \times 10^{42}$\,erg\,s$^{-1}$ \citep{DiGesu2016}.

\begin{acknowledgements}
This project was supported by the Polish NSC grants  2012/04/A/ST9/00083 (GB and MO), UMO-2017/26/D/ST9/01178 (GB), 2016/22/E/ST9/00061 (\L S and KB), 2013/09/B/ST9/00599 (AK, DKW, JM, MJ, and SZ), 2018/29/B/ST9/02004 (SZ), and 2013/10/M/ST9/00729 and 2015/18/A/ST9/00746 (AZ). 
This research has made use of data from the OVRO 40-m monitoring program \citep{Richards2011}, 
which is supported in part by NASA grants NNX08AW31G, NNX11A043G, and NNX14AQ89G and 
NSF grants AST-0808050 and AST-1109911. This paper made use of data from the \nustar mission, 
a project led by the California Institute of Technology, managed by the Jet Propulsion Laboratory, 
funded by the National Aeronautics and Space Administration. We thank the \nustar Operations, 
Software and Calibration teams for support with the execution and analysis of these observations. 
This research made use of the \nustar Data Analysis Software (NuSTARDAS) jointly developed by 
the ASI Science Data Center (ASDC, Italy), and the California Institute of Technology (USA).
 The authors thank C.C.~Cheung and the anonymous referee for their useful comments on the manuscript and suggestions.
\end{acknowledgements}

\end{document}